\documentclass{aa}

\usepackage{graphicx}	
\usepackage{txfonts}
\usepackage{amsmath}	
\usepackage{amssymb}	
\usepackage{multicol}        
\usepackage{bm}		
\usepackage{pdflscape}	
\usepackage{color} 
\usepackage{hyperref}
\usepackage{placeins}
\usepackage{booktabs}
\usepackage{multirow}
\usepackage{csquotes}
\usepackage[
  locale=US,
  separate-uncertainty=true,
  per-mode=symbol-or-fraction,
  range-phrase=-,
]{siunitx}
\usepackage{multirow}

\DeclareSIUnit{\arbitraryunit}{a.u.}
\DeclareSIUnit{\gigabyte}{GB}
\DeclareSIUnit{\pixel}{px}
\DeclareSIUnit{\jansky}{Jy}

\makeatletter
\renewcommand*\aa@pageof{, page \thepage{} of \pageref*{LastPage}}
\makeatother

\begin{document}
\renewcommand{\figureautorefname}{Fig.}
\renewcommand{\tableautorefname}{Tab.}
\renewcommand{\sectionautorefname}{Sect.}
\renewcommand{\equationautorefname}{Eq.}

   \title{Deep learning-based radiointerferometric imaging with GAN-aided training}
   
   \author{
    K. Schmidt\inst{1}\thanks{Contact e-mail: \href{mailto:kevin3.schmidt@tu-dortmund.de}{kevin3.schmidt@tu-dortmund.de}},
    F. Geyer\inst{1}\thanks{Contact e-mail: \href{mailto:felix.geyer@tu-dortmund.de}{felix.geyer@tu-dortmund.de}},
    J. Kummer\inst{2,3},
    M. Brüggen\inst{2},
    H.~W.~Edler\inst{2},
    D. Elsässer\inst{1},
    F.~Griese\inst{3,4,5},
    A.~Poggenpohl\inst{1},
    L.~Rustige\inst{3,6}
    \and
    W. Rhode\inst{1}
    }

   \institute{Astroparticle Physics, TU Dortmund University,
              Otto-Hahn-Stra{\ss}e 4a, 44227 Dortmund, Germany
         \and
             Hamburger Sternwarte, University of Hamburg,
             Gojenbergsweg 112, 21029 Hamburg, Germany
        \and
            Center for Data and Computing in Natural Sciences (CDCS), Notkestrasse 9, 22607 Hamburg, Germany
        \and 
            Section for Biomedical Imaging, University Medical Center Hamburg-Eppendorf, D-20246 Hamburg, Germany
        \and 
            Institute for Biomedical Imaging, Hamburg University of Technology, D-21073 Hamburg, Germany
        \and 
            Deutsches Elektronen-Synchrotron DESY, Notkestra{\ss}e 85, D-22607 Hamburg, Germany
            }
   \date{Received ; accepted }

 
  \abstract
   {Radio interferometry invariably suffers from an incomplete coverage of the spatial Fourier space, which leads to imaging artifacts. The current state-of-the-art technique is to create an image by Fourier-transforming the incomplete visibility data and to clean the systematic effects originating from incomplete data in Fourier space. Previously, we have shown how super-resolution methods based on convolutional neural networks can reconstruct sparse visibility data. 
   }
   {Our previous work has suffered from a low realism of the training data. The aim of this work is to build a whole simulation chain for realistic radio sources that then leads to a vastly improved neural net for the reconstruction of missing visibilities. This method offers considerable improvements in terms of speed, automatization and reproducibility over the standard techniques.}
   {Here we generate large amounts of training data by creating images of radio galaxies with a generative adversarial network (GAN) that has been trained on radio survey data. Then, we applied the Radio Interferometer Measurement Equation (RIME) in order to simulate the measurement process of a radio interferometer.}
   {We show that our neural network can reconstruct faithfully images of  realistic radio galaxies. The reconstructed images agree well with the original images in terms of the source area, integrated flux density, peak flux density, and the multi-scale structural similarity index. Finally, we show how the neural net can be adapted to estimate the uncertainties in the imaging process.}
   {}


   \titlerunning{Deep learning-based radiointerferometric imaging with GAN-aided training}
   \authorrunning{Schmidt et al.}
   \maketitle

\section{Introduction}
In modern astronomy, radio interferometry plays a unique role as it acquires the highest resolutions of astrophysical sources.
This resolution comes at the cost of low data coverage, which generates artifacts in the resulting source images.
Therefore, the cleaning of these artifacts is an unavoidable data processing tasks when dealing with radio interferometric data.

State-of-the-art radio interferometers, such as the LOw-Frequency ARray (LOFAR) \citep{lofar}, record terabytes of data per day.
This high data rate is expected to increase substantially for the Square Kilometre Array (SKA) \citep{ska}.
In order to analyze the large amounts of data on reasonable time scales, it is inevitable to adapt existing analysis strategies.
Here, machine learning techniques are a promising way to speed up and simplify existing imaging pipelines.
Especially deep learning, which is known for its fast execution times on image data, can accelerate the analysis of large data volumes.

In recent months, an increasing number of deep learning techniques have been applied to radio interferometer data.
Application examples include source detection techniques for three-dimensional Atacama Large Millimeter/submillimeter Array (ALMA) data \citep{alma_dl} and the resolution improvements of the Event Horizon Telescope's (EHT) image of the black hole in M87 using principal-component interferometric modeling (PRIMO) \citep{primo_m87}.

\begin{figure*}
    \centering
    \includegraphics[width=\hsize]{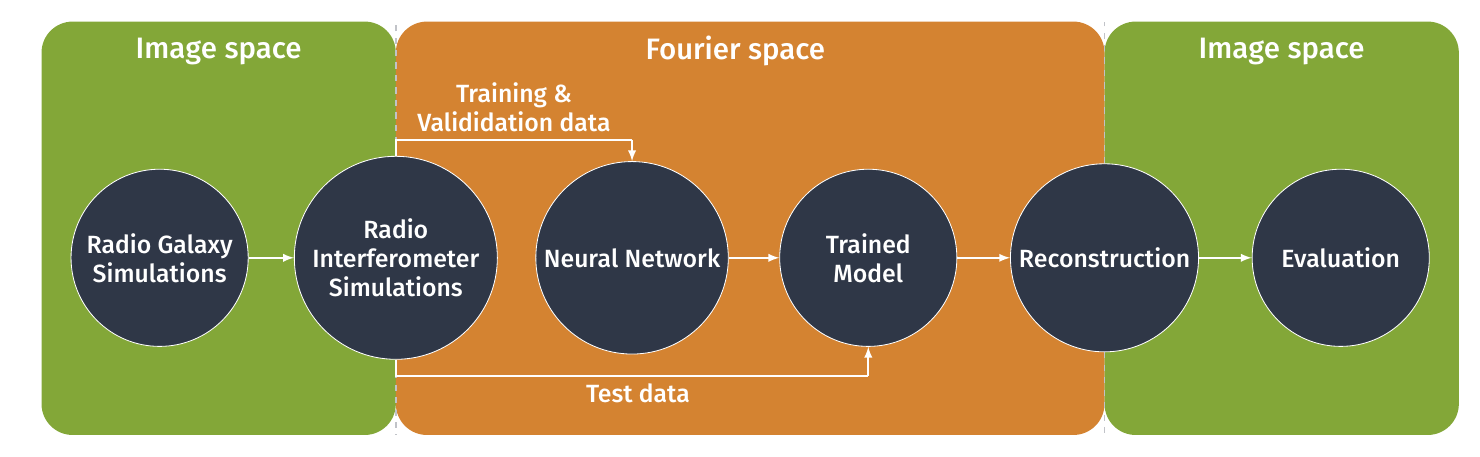}
    \caption{Analysis chain used in this work.
    The analysis chain handles data in image space (green) and Fourier space (orange).
    It is structured in three parts: simulations, neural network model, and evaluation.}
    \label{fig:analysis-chain}
\end{figure*}

We have developed a deep learning-based imaging approach for radio interferometric data published as the Python package \texttt{radionets} \citep{radionets}.
Our imaging approach uses convolutional neural networks known from super-resolution applications \citep{superres} to reconstruct missing visibility data from sparse radio interferometer layouts.
Consequently, our deep learning approach is no classical cleaning, as applied in the CLEAN algorithm \citep{clean}. Instead, we perform the data reconstruction directly in Fourier space.
The results presented in this paper build on our previous work \citep{Schmidt_2022}.
Here, we describe a new method to simulate images with generative adversarial networks (GANs), introduce an enhanced radio interferometer simulation chain based on the radio interferometer measurement equation (RIME) and improved diagnostics.

\autoref{fig:analysis-chain} provides an overview of all parts of our analysis chain. The training of deep learning models requires large amounts of training data which we provide by simulating radio galaxies in image space.
We utilize the GAN that was developed and trained by \citet{Kummer_2022,Rustige_2022} to create source distributions based on the Faint Images of the Radio Sky at Twenty-Centimeters (FIRST) survey \citep{FIRST_1995}.
In a second step, we employ the RIME framework to simulate radio interferometric observations of these sources.
The RIME simulation routine is made available in the python package \texttt{pyvisgen} \citep{pyvisgen}.
During this simulation step, the data is converted from image space into Fourier space, as radio interferometers measure complex visibilities.
We create training and validation data sets consisting of incomplete $uv$ planes.
Then, we train our neural network model to reconstruct the missing visibility data.
In the following step, dedicated test data sets are used to evaluate the reconstruction ability of the trained model.
By applying the inverse Fourier transformation to the reconstructed visibility data, we recover the source distribution in image space.
Thus we can compare the computed reconstruction to the simulated source distribution.
Furthermore, we evaluate the source reconstructions with different metrics, such as the area ratio or the intensity ratio between simulation and reconstruction.
A comparison with the established imaging software \texttt{WSCLEAN} \citep{wsclean} is performed to compare application times and reconstruction quality.

In \autoref{sec:gan-simulations}, we introduce the framework created by \citet{Kummer_2022,Rustige_2022} which provides the radio galaxy simulations. 
The additional simulation techniques used for creating visibilities are described in \autoref{sec:rime}.
\autoref{sec:dl_model} lists the changes done to our neural network model in comparison to \cite{Schmidt_2022} and the hyperparameters set for the training process.
In \autoref{sec:evaluation}, we explain our evaluation techniques.
Our upcoming projects and ideas are presented in \autoref{sec:future-work}.
In \autoref{sec:conclusions}, we summarize our results and conclude.
\section{GAN simulations}
\label{sec:gan-simulations}

\begin{figure}
    \centering
    \includegraphics[width=.49\textwidth]{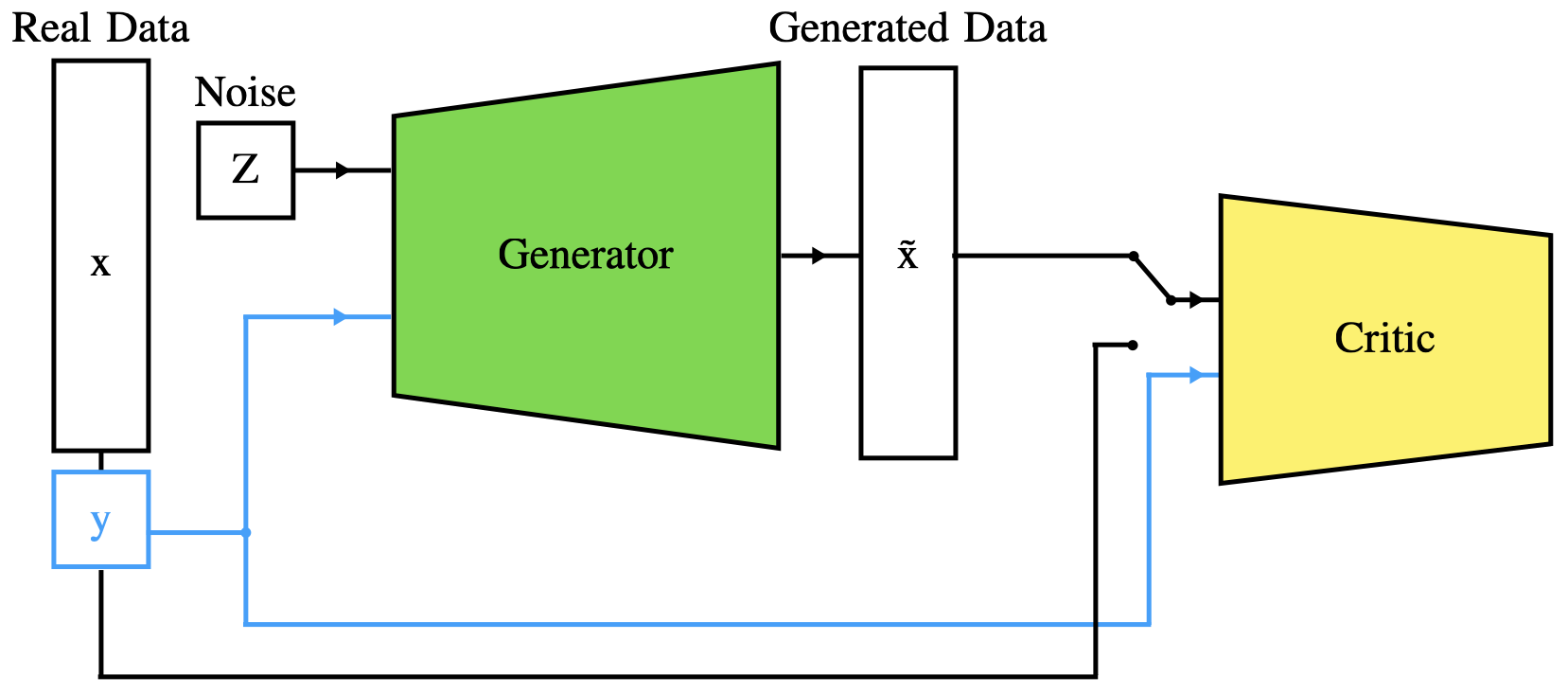}
    \caption{Schematics of the wGAN architecture reproduced from \citet{Rustige_2022}, where $y$ denotes the class label of real $x$ or generated images $\tilde{x}$}
    \label{fig:wgan_schematic}
\end{figure}

We use the framework presented in \citet{Kummer_2022,Rustige_2022} to improve the quality and authenticity of the simulated radio sources used as training data. The simulation technique is based on a generative model. Such models are able to learn the underlying statistical distribution of the data and can be used for simulations by sampling from this distribution. In our neural network-based setup a Generative Adversarial Network \citep{goodfellow2014generative,salimans2016improved} is trained in a supervised way on observations from the FIRST survey \citep{FIRST_1995}. The data was recorded by the Very Large Array (VLA) in New Mexico with a resolution of $5"$. The training data set of this generative model is presented in \citet{Griese_2023} and publicly available \citep{Griese_zenodo_2022}. The two neural networks in the standard GAN setup are called generator and discriminator. The generator generates fake images from a noise vector and the discriminator discriminates between real and fake images. Both are trained simultaneously in a two-player minimax game. Eventually, the generator learns to generate images, which are hardly distinguishable from the training images. The used setup is shown in \autoref{fig:wgan_schematic}.

Here, we employ an advanced version of the standard GAN setup, namely a Wasserstein GAN (wGAN). The  Wasserstein distance is used in the  main term of the loss function. The discriminator is replaced by a critic which is used to estimate the Wasserstein distance between real and generated images \citep{arjovsky2017wasserstein}. Additionally, both neural networks of our wGAN setup are conditioned on the morphological class label. Hence, the generator can be used to simulate images of specific classes.
For further development of our image reconstruction technique, we employ the generator of the wGAN trained for augmenting classifier training in \citet{Rustige_2022}. This setup can be used to simulate radio sources of four different morphological classes, namely FRI and FRII, which are used for this paper, and \enquote{bent} and \enquote{compact}. As the model was trained on FIRST images, the generated images have similar properties as the training set of the wGAN. For details of the model training we refer to \citet{Rustige_2022}. 

We simulate \num{30000} sources of the class FRI and \num{30000} sources of the class FRII to construct a data set which we split into a \num{5}:\num{1} ratio between training and validation data. Additionally, we simulate \num{10000} extra sources for the creation of a test data set, which are split into \num{5000} of the class FRI and \num{5000} of the class FRII. The morphological classes of the Fanaroff-–Riley classification scheme are distinguished based on the peak location and properties of the radio emission and are well-established in radio astronomy \citep{FR_1974}.
Note, for the simulation of each class a different generator model is used. Consequently, we make sure that the model is optimally chosen for the given class. 
The noise in the training data is reduced by setting all pixel values below three times the local root mean square (RMS) noise to the value of this threshold. Subsequently, the pixel values were rescaled to the range between -1 and 1 to represent floating point grayscale images. The generated images reproduce the preprocessed noise level of the training data inherently. The output size of the images is variable, but in the following we set it to $\SI{128}{\pixel} \times \SI{128}{\pixel}$. 
However, the generated images require some postprocessing: Gaussian smoothing is applied to the images due to the fact that strong variations exist between neighboring pixels around the sources, i.e. we observe low-intensity pixels directly next to high-intensity pixels. The width of the Gaussian has to be chosen such that the peaks of the main sources are not smeared out and substructures in the images are not eradicated. We decided to use a width of $\sigma=\SI{0.75}{\pixel}$, which best fulfills the previously mentioned criteria. A comparison of different $\sigma$ values for the same exemplary, wGAN-generated source is shown in \autoref{fig:gaussian_filter}. 
After that, the sources are scaled between \SI{1}{\milli\jansky} and \SI{300}{\milli\jansky}. Thus, we ensure that the signal-to-noise ratios (SNRs) range between \num{1} and \num{100}.
SNRs are defined with the help of \texttt{WSCLEAN}.
We perform a cleaning until we reach the threshold of \SI{5}{\sigma}, see \texttt{auto-mask} option of \texttt{WSCLEAN}.
Afterward, we compare the maximum of the clean image with the standard deviation of the residual image.
Tests with different simulation options revealed that flux densities between \SI{1}{\milli\jansky} and \SI{300}{\milli\jansky} lead to SNRs between \num{1} and \num{100} in the simulated observations for our chosen setup.
A detailed analysis for data with different SNRs and additional information about the SNR calculation can be found in \autoref{sec:comparison_wsclean}.

\section{RIME simulations}
\label{sec:rime}

The key to a deep learning-based analysis of radio observations are the training data sets for the deep learning models.
An accurate simulation chain is essential for the training data to have the same properties as actual measurements.
In addition to the source simulations, see \autoref{sec:gan-simulations}, the simulation of the radio interferometer plays a central role and forms the second large block in our analysis chain presented in \autoref{fig:analysis-chain}.
To describe the signal path from the sources towards measured data, we use the Radio Interferometer Measurement Equation (RIME) \citep{rime}.
The RIME utilizes Jones matrices \citep{jones} for the processing of the source signals.
For each corruption effect a new Jones matrix is defined.
The GAN-simulated sources described in \autoref{sec:gan-simulations} are the input for the RIME calculations.
Please note that matrices are marked with bold letters in the following.

As we focus on imaging tasks, we assume calibrated data for our simulations, so no direction-dependent effects are taken into account.
Hence, only the geometric signal delay between the antennas and the characteristic of the individual antenna responses are considered.
Thus, the RIME to calculate the complex visibility measured by the antenna pair $pq$ at time $t$ results in
\begin{equation}
    \mathbf{V}_{\mathrm{pq}}(l, m) = \sum_{l, m} \mathbf{E}_{\mathrm{p}}(l, m) \, \mathbf{K}_{\mathrm{p}}(l, m) \, \mathbf{B}(l, m) \, \mathbf{K}_{\mathrm{q}}^{\mathrm{H}}(l, m) \, \mathbf{E}_{\mathrm{q}}^{\mathrm{H}}(l, m).
\end{equation}
Here, $\mathbf{B}(l, m)$ describes the source's brightness distribution given in the direction cosines $l, m$.
At this stage, we do not simulate polarization data.
Generated visibilities correspond to full intensity data, known as Jones I.
Furthermore, $\mathbf{K}(l, m)$ describes the phase delay kernel defined as
\begin{align}
    \mathbf{K}(l, m) &= \exp \left(-2\pi \cdot i \cdot \left(ul + vm\right)\right).
\end{align}
The phase delay kernel takes into account the signal's propagation lengths to the different telescopes inside the interferometer array.
While $(l, m)$ are used for describing the source distribution, the antenna positions are given in the direction cosines $(u, v)$.
The effects of the phase delay kernel $\mathbf{K}(l, m)$ for different baseline lengths is illustrated in \autoref{fig:phase_delay}.
The characteristics of the VLA's \SI{25}{\meter} antennas are encoded in $\mathbf{E}(l, m)$, which is defined as
\begin{align}
    &\mathbf{E}(l, m) = \mathrm{jinc} \left(\frac{2\pi}{\lambda_{\mathrm{obs}}} d \cdot \theta_{lm}\right),\\
    &\mathrm{with}~\mathrm{jinc} = \frac{J_1(x)}{x}.
\end{align}
Again, $(u, v)$ and $(l, m)$ are the coordinate system in Fourier and image space, respectively.
Furthermore, the telescope's diameter $d$ and the angular distance between pointing position and source structure $\theta_{lm}$ are considered.
$J_1$ is the Bessel function of the first kind \citep{rime}.
The telescope response $\mathbf{E}(l, m)$ is illustrated in \autoref{fig:telescope_response}.
$\mathbf{X}_{\mathrm{q}}^{\mathrm{H}}$ corresponds to the conjugate transpose of the matrix $\mathbf{X}_{\mathrm{q}}$.
The summation over all image pixels is performed to calculate one complex visibility.

\begin{figure}
    \centering
    \includegraphics[width=\hsize]{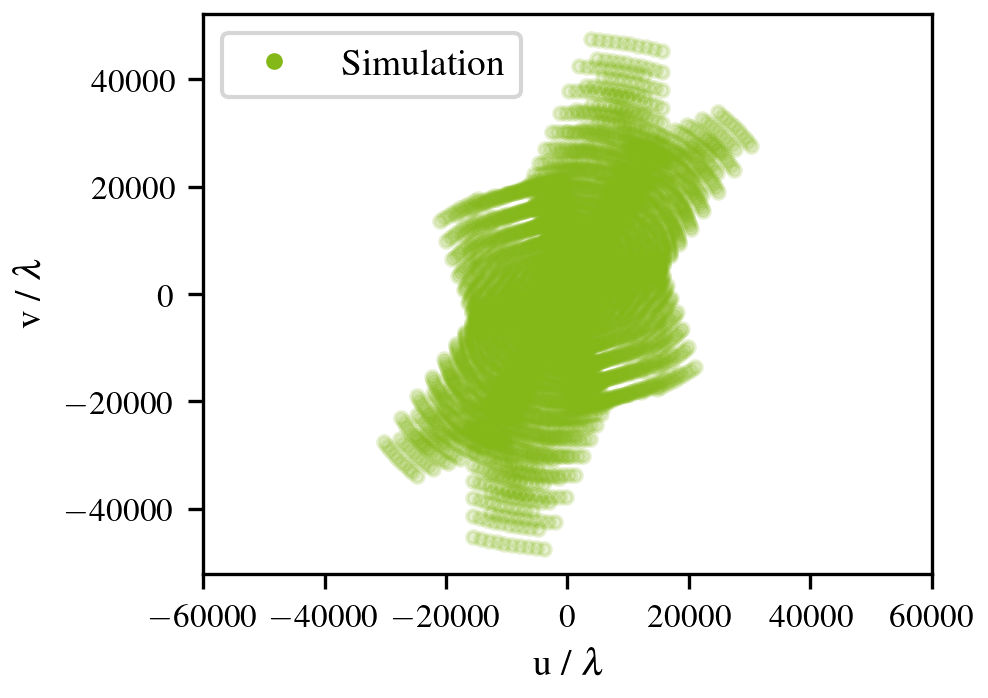}
    \caption{Exemplary $(u, v)$ coverage of a VLA simulation using the developed RIME implementation.}
    \label{fig:example-visibilities}
\end{figure}

In this analysis, we use the B-configuration layout of the VLA, which was utilized for the FIRST observations.
The antenna positions are listed in \autoref{tab:antennas-vla}.
Utilizing the RIME enables the simulation of exact $(u, v)$ coverages.
\autoref{fig:example-visibilities} shows a simulated $(u, v)$ coverage using the above RIME formalism.


One of the characteristics of the VLA is directly visible in this image:
While the center of the image is well covered, the edges are poorly sampled owing to the relatively short baselines compared for example to the VLBA.
This property complicates the reconstruction at the edge of the visibility space.

Since real observation depend on multiple observation parameters such as the correlator integration time or the number of scans per pointing, these parameters can be altered in our simulation framework.
\autoref{tab:parameters-vla} summarizes the parameters used for the simulations in this work.
In cases where multiple values are specified, these values serve as bounds of random values that are drawn within these bounds.
The parameters are altered for each new simulated observation.

\begin{table}
    \centering
    \caption{
    Observation parameters applied to simulate the data sets used in this work.
    In cases where multiple values are specified, these values serve as bounds of random values that are drawn within these bounds.
    }
    \begin{tabular}{lc}
    \toprule
    Sampling option & Value \\
    \midrule
    Image size & \SI{128}{\pixel} \\
    FOV center ra & [\SI{100}{\degree}, \SI{110}{\degree}] \\
    FOV center dec & [\SI{30}{\degree}, \SI{40}{\degree}] \\
    FOV size & 350 \\
    Corr. int. time & \SI{30}{\second} \\
    Scans & 2 \\
    Scan separation & \SI{360}{\second} \\
    Base frequency & \SI{1.365}{\giga\hertz} \\
    Bandwidths & \SI{3}{\mega\hertz} \\
    \# Bandwidths & 4 \\
    Channel offset & \SI{64}{\mega\hertz} \\
    \bottomrule
    \end{tabular}
    \label{tab:parameters-vla}
\end{table}

An additional noise corruption effect that occurs during radio interferometer observations results from the system noise.
This noise effect changes when using different bandwidths or correlator accumulation times.
The noise originating from a specific measurement process is Gaussian distributed with a standard deviation that is defined by
\begin{equation}
    \Delta S = \frac{1}{\eta_s} \frac{S_{\mathrm{SE}}}{\sqrt{2 \, \Delta\nu \, \tau_{\mathrm{acc}}}}.
\end{equation}
Here, $\eta_s$ is the system efficiency factor, $\Delta\nu$ corresponds to the observation's bandwidth, $\tau_{\mathrm{acc}}$ is the correlator accumulation time, and $S_{\mathrm{SE}}$ denotes the system equivalent flux density \citep{taylor}.
Values for $\Delta\nu$ and $\tau_{\mathrm{acc}}$ are taken from \autoref{tab:parameters-vla}, while $\eta_s = 0.93$ and $S_{\mathrm{SE}} = \SI{420}{\jansky}$ are taken from the VLA's specifications \cite{sefd}.
The noise drawn this way, which is in the order of \SI{10}{mJy}, is added separately to the real and the imaginary part of the simulated visibilities.
This noise handling technique enables the simulation of observed radio galaxies with different signal-to-noise ratios depending on the peak flux densities of the simulated radio galaxies.

Finally, the visibilities are saved to FITS files together with the meta data of the simulated observations.
We use our implementation of a FITS writer, which follows the specifications in AIPS Memo 117 \footnote{\url{http://www.aips.nrao.edu/aipsmemo.html}}.
Saving the simulations like this enables their readability and interpretability of established imaging tools like CASA \citep{casa} and \texttt{WSCLEAN} \citep{wsclean}.

Both established imaging software and our deep learning-based approach require input data on a regular grid or two-dimensional image data.
Therefore, the simulated visibilities have to be gridded.
In the context of the RIME simulations, we have implemented our own gridder which is slightly different from the established approach.
As with the gridder implemented in \texttt{WSCLEAN}, the grid can be defined with the help of the selected pixel size and number of pixels.
Afterwards our implementation does not perform a convolutional gridding, as \texttt{WSCLEAN} does, but we apply a two-dimensional histogram to grid the visibilities.
In this work, we choose a pixel size \SI{1.56}{\arcsecond} and an image size of $\SI{128}{\pixel} \times \SI{128}{\pixel}$, which corresponds to a field of view of \SI{200}{\arcsecond}.

Together with our college Stefan Fröse, we have published our simulation routine as an open-source Python package called \texttt{pyvisgen} \citep{pyvisgen}.
With this RIME simulation framework, we have built a flexible basis for future simulations of radio interferometer observations.
Different corruption effects can be easily added.
Furthermore, our method can readily be adapted for other radio interferometer arrays by updating the interferometer characteristics and the antenna positions.
\section{Deep learning-based imaging}
\label{sec:dl_model}

In this work, we perform the imaging of radio interferometer data using deep learning techniques available inside the Python package \texttt{radionets} \citep{radionets}.
Our approach differs from the conventional CLEAN algorithm \citep{clean} in that we reconstruct missing information directly in Fourier space.
In the reconstruction process, no iterative source model is formed. 
Consequently, no convolution with the clean beam of the observation is necessary.

\begin{figure}
    \centering
    \includegraphics[width=0.42\textwidth]{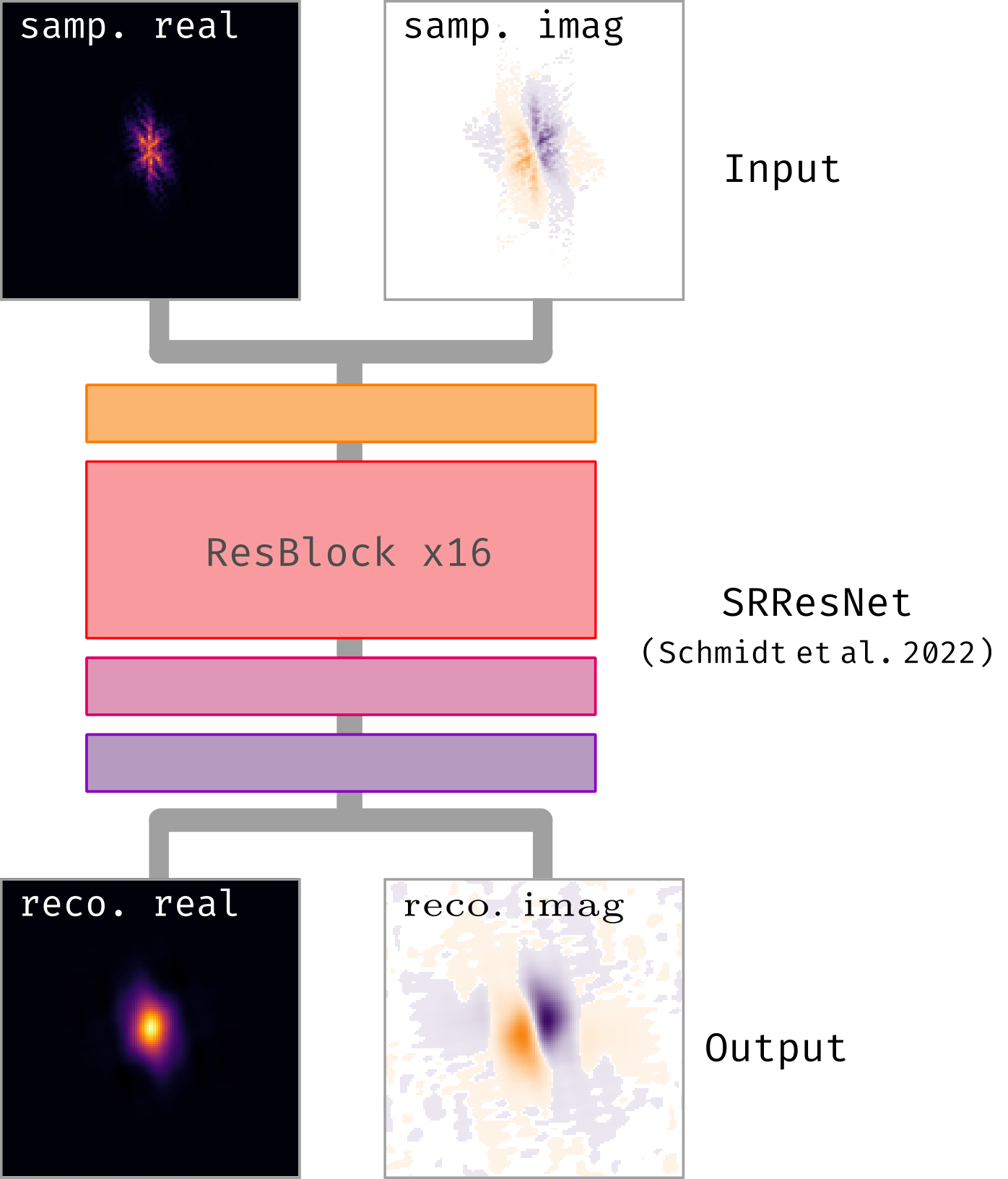}
    \caption{
    Overview of the training routine.
    Sampled real and imaginary maps generated using the RIME simulations, see \autoref{sec:rime}, serve as input for the neural network.
    We use a similar network architecture as in our previous publication \citep{Schmidt_2022}, despite an increased number of residual blocks, which was expanded from \num{8} to \num{16}.
    During training, the networks learns to reconstruct the sampled input maps.
    The output of our neural network model are the reconstructed real and imaginary maps.
    }
    \label{fig:train_overview}
\end{figure}

The reconstruction of the missing data in Fourier space is performed by the trained deep learning model.
The model exploits the data recorded during an observation and uses the contained information to estimate values for missing pixels with the help of convolutional layers \citep{Goodfellow_convs}.
We describe the underlying methodology in our previous publication \citet{Schmidt_2022}.
In the following, we discuss the changes we made to handle the simulated FIRST data.

\subsection{Model definition}

Our imaging architecture is based on residual blocks \citep{residual-nets, residual-block} integrated into a \texttt{SRResNet} architecture known from super-resolution problems \citep{superres}.
A detailed model description can be found in \cite{Schmidt_2022}.
We keep the original architecture except for increasing the number of residual blocks to sixteen. The reason for this is that as the complexity of the input images increases, as described in the previous sections, the depth and complexity of the network must also increase in order to obtain meaningful reconstructions.

\subsection{Model training}

Since our deep learning-based imaging is performed directly in Fourier space,
the deep learning model training is done in Fourier space as well.
The simulated visibilities, described in \autoref{sec:rime}, are used as input.
During the training process, the model learns to reconstruct missing information and generates complete real and imaginary maps as output.
\autoref{fig:train_overview} gives an overview of said training routine.

For model training, we use a data set consisting of \num{70000} real and imaginary images. \num{50000} of these images are used for the training of the neural network.
Additional \num{10000} validation maps help to check for over-training.
In \autoref{sec:evaluation}, the remaining \num{10000} images are used for evaluating the trained model.
In one training epoch all training and validation maps are passed through the network.
The loss function that quantifies the difference between the network's prediction and the simulated truth is chosen to be a split L1 loss \eqref{eq:l1}, which is defined as:

\begin{align}
    \mathrm{Loss} &= \mathrm{L1}\left(x_{\mathrm{real}}, y_{\mathrm{real}}\right) + \mathrm{L1}\left(x_{\mathrm{imag}}, y_{\mathrm{imag}}\right),  \label{eq:l1} \\
    \mathrm{with} &~~\mathrm{L1}(x, y) = \left|~x - y~\right|,
\end{align}
where $x$ is the predicted output from the network and $y$ is the true distribution.

We used the ADAM optimizer \citep{adam} to update the weights during training, which outperformed stochastic gradient decent (SGD) \citep{sgd} in convergence time at the cost of more readily learned parameters.
We trained the neural network model for \num{400} epochs with a batch size of \num{100}.
An adaptive learning rate is applied, starting with a learning rate of $2\cdot10^{-4}$ which peaks after 150 epochs at $1\cdot10^{-3}$.
Afterward, the learning rate steadily decrease to $6\cdot10^{-4}$ until the end of the training.
The duration of one epoch is \SI{185}{\second}, which resulted in a total training time of $\approx$ \SI{20.5}{\hour}, on the computing specifications summarized in \autoref{tab:computer-specifications}.
The reconstruction times of the neural network model are of the order of milliseconds.
For a $\SI{128}{\pixel} \times \SI{128}{\pixel}$ image the pure reconstruction time of the neural network is \SI{6.4\pm0.1}{\milli\second} evaluated 100 times on the same image.
As our approach has no iterative routines, the reconstruction time is independent of the $(u, v)$ coverage.
A reconstruction time comparison with established imaging software is given in \autoref{sec:times}.

\begin{figure}
    \centering
    \includegraphics[width=\hsize]{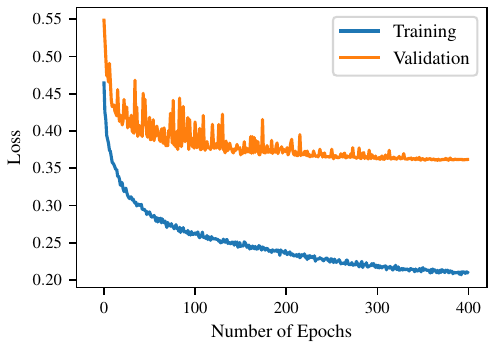}
    \caption{
    Loss curves for the training process of the network.
    Loss values as a function of epoch are shown separately for training and validation data.
    }
    \label{fig:train_loss}
\end{figure}

\autoref{fig:train_loss} visualizes the curves for the training and validation loss.
Both, training and validation losses are steadily decreasing in both cases.
A gap between training and validation loss starts to form after the first $\sim10$ epochs.
However, this gap is not a sign of over-training, as the validation loss does not start to increase in later epochs.

We were able to smooth the loss curves by implementing a normalization based on all training images.
Here, the mean of all non-zero input values is subtracted from all non-zero pixels.
In the next step, the result is divided by the standard deviation of all non-zero values.
The complete normalization is defined as
\begin{align}
    x_\text{norm} = \frac{x - \mathrm{mean}(x)}{\mathrm{std}(x)},~~\text{for}~~x\neq0.
\end{align}

\begin{figure*}
    \centering
    \includegraphics[width=\hsize]{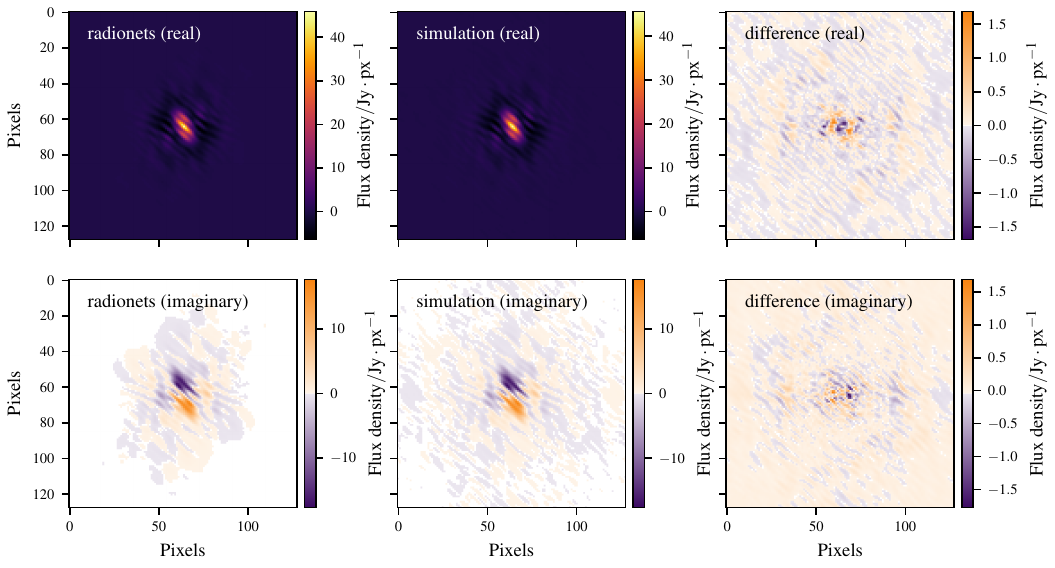}
    \caption{
    Exemplary reconstruction in Fourier space for the deep learning-based cleaning approach.
    Visualization of prediction (left), true distribution (middle) and the difference between both (right).
    Results are shown for real (top) and imaginary (bottom) part.
    }
    \label{fig:compl_reco}
\end{figure*}

\subsection{Reconstruction}

In this section, we show how the deep learning model is able to reconstruct incomplete visibility data.
\autoref{fig:compl_reco} shows the reconstructed real and imaginary maps for an exemplary test source. The top row illustrates the reconstruction for the real part, while the bottom row is dedicated to the imaginary part.

The distributions shown in \autoref{fig:compl_reco} reveal a good agreement between reconstruction and simulated truth.
Especially for the real part, all structures in the center of the image are reconstructed well. For the imaginary part, the center is also well reconstructed, while the outer parts are predicted to be a constant. Since in the true image these parts are not significant, this is not a big effect on the reconstruction. The differences for, both, real and imaginary part are an order of magnitude lower than the flux density in the original image.

In \cite{Schmidt_2022}, we
used amplitude and phase instead of real and imaginary parts of the complex visibilities.
In the case of simple Gaussian sources, using amplitude and phase instead of real and imaginary parts had the advantage that the parameter space was smaller.
However, we noticed a sharp decrease in the phase reconstructions when switching to more realistic data.
Our experiments showed that in the case of FIRST simulations the reconstruction got much better with real and imaginary maps.  \autoref{fig:imag_vs_phase} shows that the reconstructions of the imaginary part are reasonably good up to the edges, while phase map reconstructions deviate more significantly from the true values in these areas.
\section{Diagnostics}
\label{sec:evaluation}

\begin{figure*}
    \centering
    \includegraphics[width=\hsize]{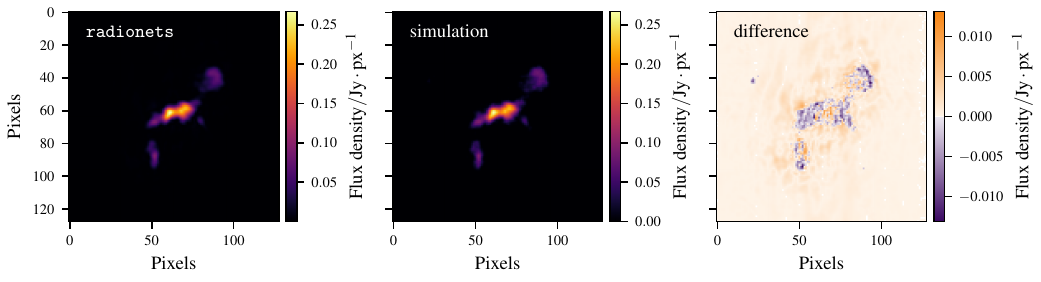}
    \caption{Image space reconstruction for the deep learning-based cleaning approach.
    Visualization of predicted source distribution (left), simulated source distribution (middle) and the difference between both (right).
    The real and imaginary visibility maps used for the inverse Fourier transformation are those shown in \autoref{fig:compl_reco}.
    }
    \label{fig:image-pred}
\end{figure*}

\begin{figure*}
    \centering
    \includegraphics[width=\hsize]{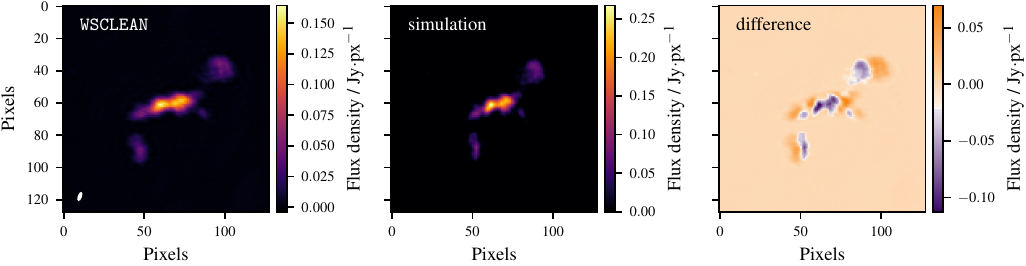}
    \caption{
    Clean image generated using \texttt{WSCLEAN}. Reconstructed clean image (left), simulated source distribution (middle), and difference between reconstruction and simulation (right). The \texttt{WSCLEAN} reconstruction was converted to units of $\mathrm{Jy}\cdot\mathrm{px}^{-1}$ for comparability with the simulations. For the clean image generated by \texttt{WSCLEAN}, the clean beam is shown in the lower left edge.
    }
    \label{fig:reco_wsclean}
\end{figure*}

In this section, we quantify the reconstruction ability of our deep learning model.
The diagnostics will be performed in image space, which means the reconstructed visibility distributions are processed by the inverse Fourier transformation before the evaluation techniques get applied on the reconstructed source distributions.
Furthermore, we compare our deep learning-based approach with the
Cotton-Schwab Clean algorithm \citep{schwab} implemented in
the established imaging software \texttt{WSCLEAN} \citep{wsclean}.
\autoref{tab:wsclean-params} summarizes the cleaning parameters, which were used for all cleaning tasks with \texttt{WSCLEAN} in this work. Due to the increasing complexity and computation times, we restricted ourselves to a small number of tuneable parameters. Therefore, we omit e.g. the multi-scale cleaning and use the conservative Cotton-Schwab clean algorithm.
For both \texttt{radionets} and \texttt{WSCLEAN}, the goal is that reconstructed and simulated radio galaxies do not deviate.

\begin{table}
    \centering
    \caption{Overview of the parameter settings used to create the clean images using \texttt{WSCLEAN}.}
    \begin{tabular}{lr}
    \toprule
    Parameter & Setting \\
    \midrule
    size & \SI{128}{px} \\
    scale & \SI{1.56}{asec} \\
    mgain & \num{0.8} \\
    gain & \num{0.01} \\
    niter & \num{1000000} \\
    auto-mask & \num{3} \\
    autothresh & \num{1} \\
    weight & uniform \\
    \bottomrule
    \end{tabular}
    \label{tab:wsclean-params}
\end{table}

\subsection{Comparison with \texttt{WSCLEAN}}
\label{sec:comparison_wsclean}

As a first evaluation, we perform a visual analysis of the inverse Fourier transformed visibility maps shown in \autoref{fig:compl_reco}. The corresponding source distributions in image space are shown in \autoref{fig:image-pred}.
In particular, we show the deep learning-based source reconstruction, the simulated source distribution and their difference. 
The mean values of the differences are an order of magnitude smaller than the peak flux densities, meaning that simulation and prediction only differ slightly.
\autoref{fig:reco_wsclean} shows the reconstruction for the same source using \texttt{WSCLEAN}.
The comparison between the reconstructed and the simulated source distribution reveals a smeared-out source structure in the case of \texttt{WSCLEAN}.
This smearing results from the convolution with the restoring beam
inside \texttt{WSCLEAN}'s cleaning routine.

To test the model's performance, we need to evaluate more than one image.
To this end, we have developed three diagnostic methods that focus on different aspects of a complete source reconstruction.
We also use an existing evaluation metric, the Multi Scale Structural Similarity Index Measure (MS-SSIM), to compare the similarity between reconstructed and simulate source images.
All four evaluation methods were applied to a distinctive test data set containing \num{10000} images.
The test images are the same for the \texttt{radionets} and the \texttt{WSCLEAN} evaluation.

\begin{figure*}
\centering
\begin{minipage}[t]{0.48\textwidth}
    \centering
    \includegraphics[width=\hsize]{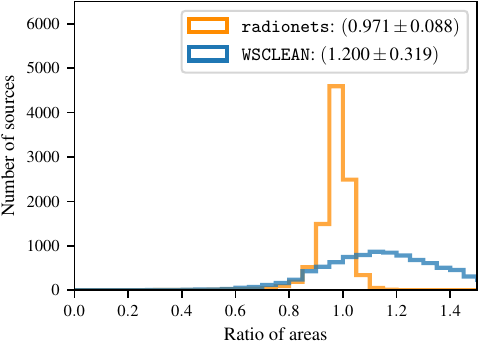}
    \caption{Ratio of the true and the predicted areas for the \texttt{radionets} and \texttt{WSCLEAN} reconstructions. For both reconstruction approaches the mean and standard deviation is given. In the case of \texttt{WSCLEAN}, the area ratio is corrected for the beam smearing.}
    \label{fig:hist-area}
\end{minipage}
\hfill%
\begin{minipage}[t]{0.48\textwidth}
\centering
\includegraphics[width=\hsize]{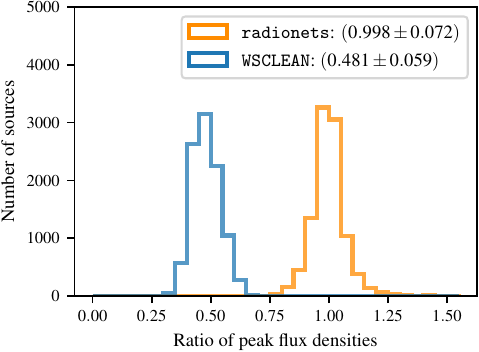}
\caption{Ratio of the true and the predicted peak flux densities for the \texttt{radionets} and \texttt{WSCLEAN} reconstructions. For both reconstruction approaches the mean and standard deviation is given. In the case of \texttt{WSCLEAN}, the flux density is converted from units of \si{Jy\cdot beam^{-1}} to \si{Jy\cdot px^{-1}}. 
The obvious difference between the distributions originates from different effective resolutions that the methods can achieve.}
\label{fig:hist-peak}
\end{minipage}\\[2em]
\begin{minipage}[t]{0.48\textwidth}
    \centering
    \includegraphics[width=\hsize]{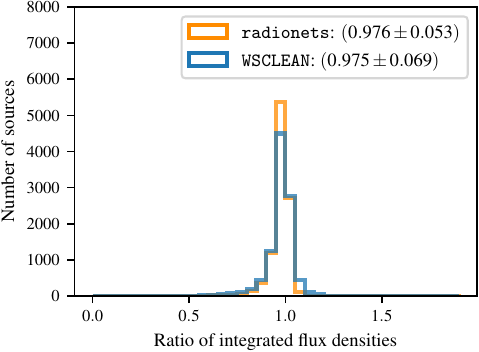}
    \caption{Ratio of the true and the predicted integrated flux densities for the \texttt{radionets} and \texttt{WSCLEAN} reconstructions. For both reconstruction approaches the mean and standard deviation is given. In the case of \texttt{WSCLEAN}, the flux density is converted from units of \si{Jy\cdot beam^{-1}} to \si{Jy\cdot px^{-1}}.}
    \label{fig:hist-sum}
\end{minipage}
\hfill
\begin{minipage}[t]{0.48\textwidth}
    \centering
    \includegraphics[width=\hsize]{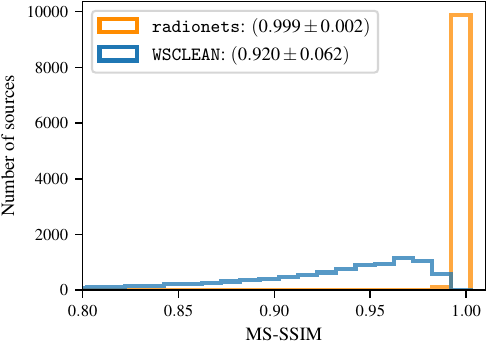}
    \caption{Multi Scale Structural Similarity Index Measure for the \texttt{radionets} and \texttt{WSCLEAN} reconstructions. For both reconstruction approaches the mean and standard deviation is given. For a fair comparison of the source structures, the true source distributions are smeared out with the clean beam calculated by \texttt{WSCLEAN}.}
    \label{fig:hist-msssim}
\end{minipage}
\end{figure*}

First, we compare the reconstructed and simulated source areas. 
The necessary cut to distinguish source and background structures is \SI{10}{\percent} of the peak flux density.
Then, the area of every source component in the image is computed and summed up for both predicted and simulated source distribution. This is done by obtaining the contour levels using \texttt{matplotlib} and and relating them to the enclosed area using Leibniz' Sector formula. A more detailed explanation can be found in our previous work \citep{Schmidt_2022}.
\autoref{fig:hist-area} shows the area ratios for the \texttt{radionets} and the \texttt{WSCLEAN} reconstructions.
The optimal value of one is reached in cases where the source areas of, both, prediction and simulation are exactly the same.
Ratios below one indicate an underestimated source area, while ratios above one denote overestimated areas. 
As shown in the histogram, for the majority of test sources reconstructed and simulated source areas match well in the case of \texttt{radionets}, which is confirmed by the mean ratio of \num{0.971} and the standard deviation of \num{0.088}.
In the case of \texttt{WSCLEAN}, the source area is overestimated around \SI{20}{\percent} for most of the sources.
The reason for this overestimate is the beam smearing in \texttt{WSCLEAN}'s reconstruction routine.
After creating a point source model, the built model is 
convolved with the clean beam of the observation.
This procedure spreads the source emission to a larger area of the sky.
For a fair comparison between the \texttt{radionets} and \texttt{WSCLEAN} reconstruction, we take into account the smearing during our comparison with the simulated images.
We use the clean beam computed by \texttt{WSCLEAN} to smear the simulated source distribution before we calculate the ratios for \texttt{WSCLEAN}.

The second method examines the peak flux densities of the predicted and simulated source distributions.
Again, the above threshold is applied to define the source.
Then, the ratio between predicted and simulated peak flux densities is calculated.
\autoref{fig:hist-peak} shows the values for the \texttt{radionets} and \texttt{WSCLEAN} reconstructions.
The optimal ratio is again one, while ratios above and below indicate over- and underestimates, respectively.
The distribution peaks around the 1 with a mean ratio of 0.998 and a standard deviation of 0.072 for \texttt{radionets}.
This underlines that the brightest spot, e.g. the core of the source, is reconstructed well.
Again, the comparison with \texttt{WSCLEAN} is a bit more complicated.
Beam smearing effects have to be taken into account before the peak flux density comparison with the simulated source distribution is possible.
While the simulated flux densities are in units of $\mathrm{Jy}\cdot\mathrm{px}^{-1}$, the \texttt{WSCLEAN} reconstructions are in units of $\mathrm{Jy}\cdot\mathrm{beam}^{-1}$.
We compute the beam area, divide it by the pixel area, and use the scaling factor to scale the flux densities of the \texttt{WSCLEAN} images so that they match the simulations.
For the beam area calculation, we use the expression

\begin{align}
    A_\text{beam} = \frac{2\,\pi \, b_\text{min} b_\text{maj}}{8 \, \ln(2)},
\end{align}
where $b_\text{min}$ is the full width at half maximum (FWHM) of the beam's minor axis and $b_\text{maj}$ the FWHM of the beam's major axis.
Even though the images have the same flux density units, the peak flux density of the \texttt{WSCLEAN} reconstruction is underestimated by a factor of two with a mean ratio of \num{0.481} and a standard deviation of {0.059}.
The beam smears the initial peak flux to neighboring pixels, which causes a decrease of the peak flux density in this individual pixel. 
Another effect to take into account is that the effective resolution for \texttt{radionets} is higher than for \texttt{WSCLEAN}, as can be seen in the right images of \autoref{fig:image-pred} and \autoref{fig:reco_wsclean}. Thus, the non-correctable beam smearing and the difference in resolution lead to the expected lower peak flux densities for \texttt{WSCLEAN}.

For the third diagnostic method, we sum up all flux densities inside of the \SI{10}{\percent} peak flux density threshold to calculate the integrated flux density.
Then, we compute the ratio between the values for reconstruction and simulation.
\autoref{fig:hist-sum} shows the resulting ratios for \texttt{WSCLEAN} and \texttt{radionets}.
For both reconstruction techniques, the integrated peak flux densities match well for the complete test data set.
The good agreement is apparent in the mean and standard deviation of \num{0.976\pm0.053} for \texttt{radionets} and \num{0.975\pm0.069} for \texttt{WSCLEAN}.
In the case of the integrated flux density, the beam smearing of \texttt{WSCLEAN} plays no significant role anymore after transforming the images from units of $\mathrm{Jy}\cdot\mathrm{beam}^{-1}$ to units of $\mathrm{Jy}\cdot\mathrm{px}^{-1}$.


\begin{figure*}
\centering
\begin{minipage}[t]{0.48\textwidth}
    \centering
    \includegraphics[width=\hsize]{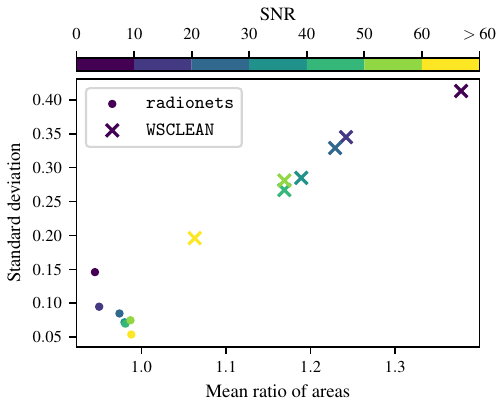}
    \caption{
    Means and standard deviations of the area ratios from \autoref{fig:hist-area} split into the different SNRs of the input visibility maps.
    Results are shown for the \texttt{radionets} and \texttt{WSCLEAN} reconstructions. 
    Both reconstruction techniques converge to the optimal mean ratio of one and standard deviation of zero for input data with higher SNR.
    While \texttt{radionets} reconstructions underestimate the source areas for smaller SNRs, \texttt{WSCLEAN} overestimates the source areas for these cases.
    }
    \label{fig:mean-std-area}
\end{minipage}
\hfill%
\begin{minipage}[t]{0.48\textwidth}
\centering
\includegraphics[width=\hsize]{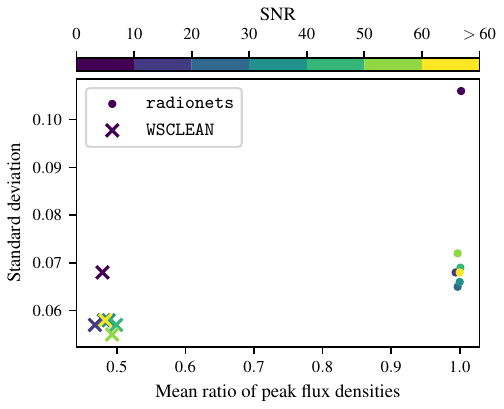}
\caption{
    Means and standard deviations of the peak flux density ratios from \autoref{fig:hist-peak} split into the different SNRs of the input visibility maps.
    Results are shown for \texttt{radionets} and \texttt{WSCLEAN} reconstructions.
    Mean ratios are not influenced by the SNRs.
    Standard deviations slightly improve with higher SNRs.
    Peak flux densities for \texttt{WSCLEAN} are underestimated, as the exact correction of beam smearing to neighboring pixels is not possible.
    }
\label{fig:mean-std-peak}
\end{minipage}\\[2em]
\begin{minipage}[t]{0.48\textwidth}
    \centering
    \includegraphics[width=\hsize]{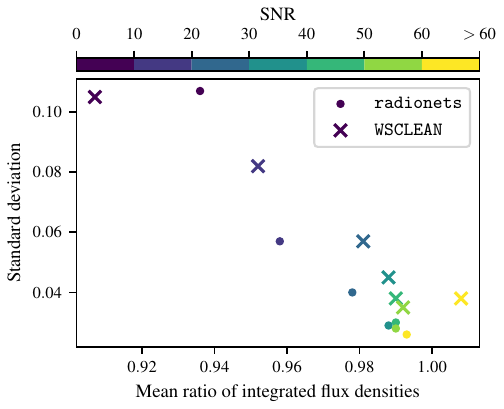}
    \caption{
    Means and standard deviations of the integrated flux density ratios from \autoref{fig:hist-sum} split into the different SNRs of the input visibility maps.
    Results are shown for \texttt{radionets} and \texttt{WSCLEAN} reconstructions.
    For both methods, the input data with the highest SNRs produce the best mean values and standard deviations.
    }
    \label{fig:mean-std-sum}
\end{minipage}
\hfill
\begin{minipage}[t]{0.48\textwidth}
    \centering
    \includegraphics[width=\hsize]{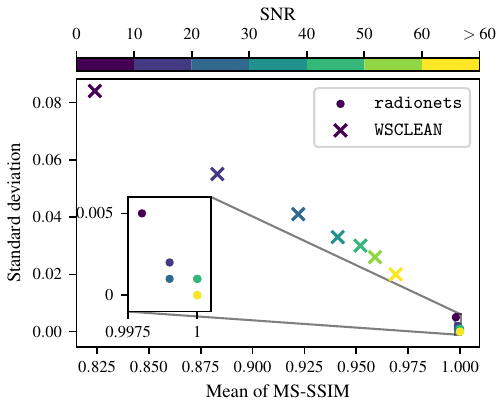}
    \caption{
    Means and standard deviations of the multi scale structural similarity indices (MS-SSiMs) from \autoref{fig:hist-msssim} split into the different SNRs of the input visibility maps.
    Results are shown for \texttt{radionets} and \texttt{WSCLEAN} reconstructions.
    For \texttt{radionets} the MS-SSIM does not significantly change for different SNRs, while the results improve with higher SNRs in the case of \texttt{WSCLEAN}.
    }
    \label{fig:mean-std-msssim}
\end{minipage}
\end{figure*}

For a more image-based comparison of the simulated and the predicted source distributions, we use a criterion commonly used in the field of computer vision, the \textbf{M}ulti \textbf{S}cale \textbf{S}tructural \textbf{S}imilarity \textbf{I}ndex \textbf{M}easure (MS-SSIM) \citep{ms-ssim}. 
This metric is split into three distinctive parts, which are luminance, contrast and structure:

\begin{align}
    l(x, y) &= \frac{2 \mu_x \mu_y + C_1}{\mu_x^2 + \mu_y^2 + C_1} \label{eq:luminance}
\end{align}
\begin{align}
    c(x, y) &= \frac{2 \sigma_x \sigma_y + C_1}{\sigma_x^2 + \sigma_y^2 + C_1} \label{eq:contrast}
\end{align}
\begin{align}
    s(x, y) &= \frac{\sigma_{xy} + C_3}{\sigma_x \sigma_y + C_3} \label{eq:structure}
\end{align}
with $\mu$ as the mean, $\sigma$ as the standard deviation, $\sigma_{xy}$ as the covariance, $x$ and $y$ as the images to be compared and

\begin{equation}
    C_1 = (K_1\,L)^2, C_2 = (K_2 \, L)^2 \ \text{and} \ C_3 = C_2/2 \, .
\end{equation}
Here, $L$ refers to the dynamic range of the pixel values and $K_1 \ll 1$ and $K_2 \ll 1$ to small scalar constants that are used to correct for numerical instabilities in the denominator. The metric is then a combination of those three components \eqref{eq:luminance} to \eqref{eq:structure}:

\begin{equation}
    \mathrm{SSIM}(x, y) = \left[l_M(x, y)\right]^{\alpha_M} \prod_{j=1}^M \left[c_j(x, y)\right]^{\beta_j} \left[s_j(x, y)\right]^{\gamma_j} \, .
\end{equation}
This formula is an improvement over the predecessor because it opens up the possibility to use image details at different resolutions. The optimal value is 1, which can only be achieved if $x=y$.

As in the previous three evaluation methods, we reconstructed \num{10000} test sources with \texttt{WSCLEAN} and \texttt{radionets}.
Then, we computed the MS-SSIM between the reconstructed and simulated source distributions, see  
\autoref{fig:hist-msssim}.
In the case of \texttt{radionets}, the distribution peaks around the optimal value of one.
Further evidence for the good performance is the fact that the mean is \num{0.999} and the standard deviation is \num{0.002}.
In terms of computer-vision and image reconstruction tasks, the predicted and simulated source distributions are very similar with very few outliers.
This proves the robustness of our approach to model, both, noisy input images and more complex source structures, which is a major improvement over the Gaussian sources from our previous work. For \texttt{WSCLEAN}, the results are worse with a mean and standard deviation of \num{0.920(62)}. 
The effect is related to the beam smearing inside \texttt{WSCLEAN}'s cleaning routine.
The deviation of the MS-SSIM is consistent with the distributions shown in \autoref{fig:hist-area}.
Even though we apply the same clean beam to the simulated sources before calculating the MS-SSIM, the beam smearing effects cannot be completely corrected, which results in an source area overestimation around \SI{20}{\percent}.
When computing the mean structural similarities, this area over-estimations lead to deviations from the optimal value of one in the case of \texttt{WSCLEAN}. All mean values and standard deviations from the four methods are summarized in \autoref{tab:overview_eval}.

\begin{table}
    \centering
    \caption{Overview of the mean values and standard deviations from \autoref{fig:hist-area} to \autoref{fig:hist-msssim}.}
    \begin{tabular}{c|c|c}
    \toprule
    method & \texttt{radionets} & \texttt{WSCLEAN} \\
    \midrule
    Area &  \num{0.971(88)} & \num{1.200(319)}\\
    Peak flux &  \num{0.998(72)} & \num{0.481(59)}\\
    Integrated flux & \num{0.976(53)} & \num{0.975(69)}\\
    MS-SSIM & \num{0.999(2)} & \num{0.920(62)} \\
    \bottomrule
    \end{tabular}
    \label{tab:overview_eval}
\end{table}

In order to test the robustness of both approaches, we split the results of the  evaluation methods into the different SNRs of the input data.
The SNRs of the input data is quantified with the help of \texttt{WSCLEAN}.
We run \texttt{WSCLEAN} on the test data with the settings summarized in \autoref{tab:wsclean-params}, but the \texttt{auto-mask} option is set to \num{5}.
Afterwards, we calculate the the SNR as follows:

\begin{align}
    \mathrm{SNR} = \frac{\mathrm{max}(I_\text{clean})}{\mathrm{std}(I_\text{residual})}.
\end{align}
Here, $I_\text{clean}$ denotes the clean image and $I_\text{residual}$ represents the residual image.
The resulting SNR distribution is shown in \autoref{fig:hist_snr}.
\autoref{fig:vis_snr} shows input data with a SNR of \num{5}, \num{35}, and \num{70} for a visual analysis of the different noise levels.

We created seven SNR categories,
0 to 10, 10 to 20, and so on with the final category begin all SNR above sixty.
For the comparison between \texttt{radionets} and \texttt{WSCLEAN}, we compute the mean and standard deviation for each category and plot the standard deviation against the mean ratio.

\autoref{fig:mean-std-area} shows the means and standard deviations of the area ratios from \autoref{fig:hist-area} split into the different SNRs of the input visibility maps.
Both reconstruction techniques converge to the optimal mean ratio of one and the optimal standard deviation of zero for larger SNRs.
While \texttt{radionets} reconstructions underestimate the source areas for smaller SNRs, \texttt{WSCLEAN} overestimates the source areas for these cases.

The means and standard deviations of the peak flux density ratios from \autoref{fig:hist-peak} are shown in \autoref{fig:mean-std-peak}.
Analyzing the results for the different SNRs of the input visibilities reveals no significant improvements regarding the mean ratios.
The standard deviation is slightly improved with increasing SNR.
Peak flux densities for \texttt{WSCLEAN} are underestimated too much, as the exact correction of beam smearing to
neighboring pixels is not possible.

In \autoref{fig:mean-std-sum} the mean and standard deviations for the integrated flux density ratios are shown.
A significant improvement for means and standard deviations is evident for \texttt{radionets} and \texttt{WSCLEAN} reconstructions.
For both reconstruction techniques, the highest SNRs converge to the optimal mean of one and the optimal standard deviation of zero.

\autoref{fig:mean-std-msssim} shows the mean and standard deviation of the MS-SSIM split into the different SNRs of the input visibilities.
For the \texttt{radionets} reconstructions, the MS-SSIM does not significantly change for different SNRs, while the results improve with higher SNRs in the case of \texttt{WSCLEAN}.
The smaller MS-SSIM values for \texttt{WSCLEAN} originate from the smeared out source structures because of the clean beam of the observation.
The clean beam improves with higher SNRs.
Consequently, the smearing decreases, which results in an improved MS-SSIM.

A detailed summary of the corresponding values is given in \autoref{tab:snr} for \texttt{radionets} and in \autoref{tab:snr-wsclean} for \texttt{WSCLEAN}.

\subsection{Computing times}
\label{sec:times}

\begin{table}
    \centering
    \caption{Comparison of run-times to reconstruct simulated observations with \texttt{WSCLEAN} and \texttt{radionets}. Computing times are shown for a $\SI{128}{\pixel} \times \SI{128}{\pixel}$ image. For \texttt{radionets}, the run-times are split into pure deep learning reconstruction times (DL reco) and reconstruction times including data operations (DL reco + I/O).}
    \begin{tabular}{lc}
         & Computing times / s \\ \toprule
    \texttt{WSCLEAN}                    &  \num{4.52\pm0.23} \\
    \texttt{radionets}: DL reco + I/O   &  \num{2.01\pm0.03} \\
    \texttt{radionets}: DL reco         &  \num{0.04\pm0.01} \\ \bottomrule
    \end{tabular}
    {\parbox{3in}{\vspace{0.15cm}\footnotesize Note: The programs are run for 100 times on one image. The presented values are the mean run-times with standard deviations. Pure deep learning reconstruction times (DL reco) just include the time needed for applying the deep learning model. Evaluation times including data loading, model loading, and data saving (DL reco + I/O) are significantly longer as these data operation tasks are not yet optimized.}}
    \label{tab:run_times}
\end{table}

A major advantage of the deep learning-based analysis are the short computing times, after the deep learning model was trained.
While \texttt{WSCLEAN} requires around \SI{4.5}{\second} to process a $\SI{128}{\pixel} \times \SI{128}{\pixel}$ simulation, our \texttt{radionets} framework less than half this time with \SI{2.01}{\second} (DL reco + I/O).
This reconstruction time includes also data operations like data loading, model loading and saving the reconstruction.
Data operations are expected to be accelerated in the future, as these are not yet optimized.
The pure reconstruction times for the deep learning model (DL reco) are even faster as convolutional neural networks (CNN) are very efficient when applied to image data.
The neural network models needs around \SI{40}{\milli\second} to reconstruct a $\SI{128}{\pixel} \times \SI{128}{\pixel}$ simulation.
All computing times are summarized in \autoref{tab:run_times}.
\FloatBarrier

\section{Future Work}
\label{sec:future-work}

We have demonstrated the potential of radio interferometric imaging using deep learning techniques.
Still some work needs to be done until this technique can be applied to real data. In particular, we will focus on two issues that we will discuss in the following.

\subsection{Uncertainty estimates}

\begin{figure*}
    \centering
    \includegraphics[width=.8\hsize]{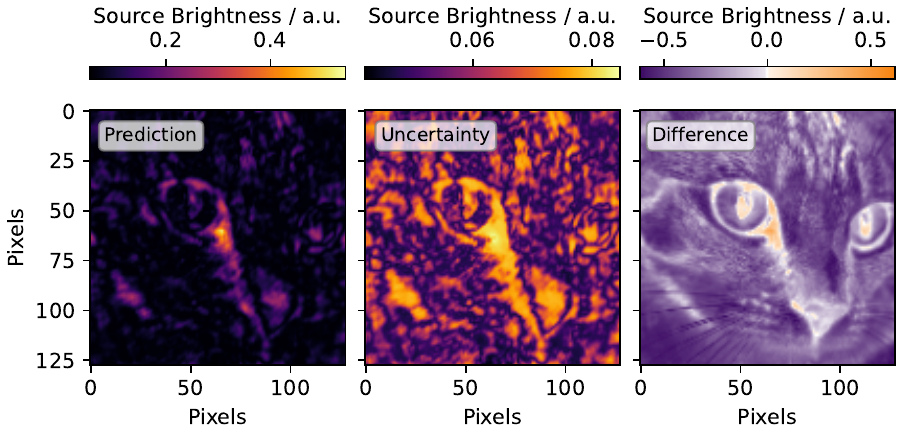}
    \caption{Overview of the prediction (left) and uncertainty (right) maps of an image of a cat. The difference between the predicted and the true image is shown on the right. The shapes of the prediction and uncertainty maps are very similar, which means that the network is uncertain with every pixel its predicted. This is to be expected, as the network is not trained on shapes like this. Image is taken from Scikit-image: Chelsea the cat by Stefan van der Walt.}
    \label{fig:cat}
\end{figure*}

\begin{figure*}
    \centering
    \includegraphics[width=.8\hsize]{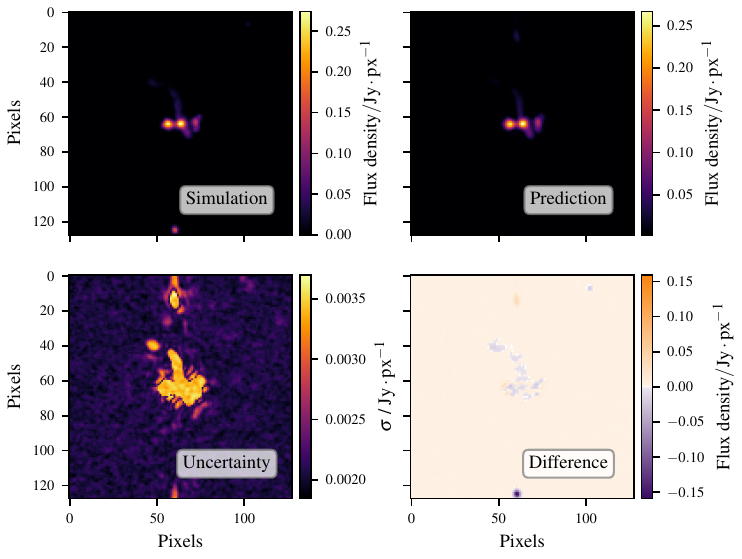}
    \caption{Reconstruction and uncertainty map of a radio galaxy from the test data set. While the shown source reconstruction (upper right) is generated by calculating the mean of the sampled reconstructions, their standard deviation gives the uncertainty map (lower left). Simulated source distribution (upper left) and prediction are in good agreement, which is reflected by their difference map (lower right).}
    \label{fig:unc-radionets}
\end{figure*}


Our neural network should not only reconstruct images, but also estimate their uncertainty.
The uncertainty can be estimated by using a negative log likelihood $l_{\beta-NLL}$ as loss function, following \cite{loss-unc} and \cite{unc-icecube}:

\begin{align}
    l(x, y) &= \frac{\log \sigma^2(x)}{2} + \frac{\left(\mu(x) - y \right)^2}{2\sigma^2(x)} \\
    l_{\beta-NLL}(x, y) &= \mathrm{stop}\left(\sigma^{2\beta}\right) l(x, y)
\end{align}
Now, a Gaussian distribution is predicted for each pixel.
The neural network model introduced in \autoref{sec:dl_model} serves to predict the mean values, while a second convolutional network is used for the prediction of the variances.
Sampling from these Gaussian distributions quantifies the uncertainties in Fourier and in image space.

Hence, we are able to estimate the reconstruction error of source types, on which the neural network has not been trained.
This can be illustrated with the example of an image of a cat, see \autoref{fig:cat} (right).
First, we simulate the observation of the cat with the mask in Fourier space used above.
Afterwards, the gridded visibilities are evaluated with our model which was trained on the FIRST simulations, as explained in \autoref{sec:gan-simulations} and \autoref{sec:rime}.
Now we are able to create the mean reconstruction as well as the corresponding uncertainty map, see \autoref{fig:cat}.
It appears that the network is able to reconstruct parts of the left eye of the cat, but not much more since it has been trained on images with a completely different shape. Consequently, the uncertainty map highlights exactly those areas that are reconstructed. This means that the network is uncertain with its prediction, which is the expected behavior. This feature could help to detect unknown source shapes. 

Additionally, we tested our uncertainty approach on source distributions from the test data set introduced in \autoref{sec:evaluation}.
\autoref{fig:unc-radionets} visualizes the reconstruction and the corresponding uncertainty map.
Simulated source distribution (upper left) and deep learning reconstruction (upper right) are in good agreement.
By computing the difference map (lower right) between simulation and prediction, two problems become apparent.
The point source below the main part of the radio source is not reconstructed in the prediction of the neural network.
Simultaneously, the prediction shows a feature above the central radio source, which is an artifact in the reconstruction.
Both features are represented in the calculated uncertainty map (lower left).
Especially for the missing point source the intensity inside the uncertainty map is not high enough to cover the difference between simulation and prediction.
For the central source structure the uncertainty map covers the deviation between simulation and prediction as it is significantly lower.
In our future work, we will focus on improving the coverage of the actual difference for the uncertainty maps.

\subsection{Wide-field and survey data}

Modern radio interferometers have wide field-of-views and record multi-frequency data.
In future work, we will focus on enhancing our simulation chain to take these specifications into account.
One approach to improve the simulations to match sky survey data is to choose larger sky sections as input for the RIME formalism.
In addition to the central main source, we will simulate other sources around the pointing center of the simulated observation.
Thus, noise from bright neighboring sources will be taken into account in the visibility data.

In order to improve the deep learning-based imaging, we plan to exploit the spectral data of different frequency channels.
Since data does not vary much between neighboring channels, the reconstruction of the first channel constitutes a good initial guess for the reconstruction of
the second channel.
Processing this information inside the deep learning model has the potential to improve and speed up the reconstruction of data cubes significantly.

Furthermore, there is still potential to improve the reconstruction by enhancing the gridder.
Currently, our gridder creates a simple two-dimensional histogram, as explained in \autoref{sec:rime}.
This approach is different from established gridders that use convolutional gridding to facilitate subsequent imaging.
Especially in connection with the large field-of-views of modern sky surveys, wide-field gridding methods will also be relevant for imaging with deep learning techniques.

\section{Conclusions}
\label{sec:conclusions}

In this paper, we have focused on improving the training data for our neural network. To this end, we have produced realistic radio galaxy simulations and modeled the measurement process of a radio interferometer.

Our image simulation is based on the GAN developed in \cite{Kummer_2022,Rustige_2022}. The GAN can produce an arbitrary number of realistic images of radio galaxies that resemble images from the FIRST survey.

These radio sources are then used as input for the RIME formalism presented in \autoref{sec:rime}. We consider the phase delay and the antenna characteristics for the simulation process and also smear the simulated measurement with noise originating from the measurement process. Various parameters for a real measurement are also taken into account, such as the coordinates of the center of the FOV randomly, as shown in \autoref{tab:parameters-vla}.

The GAN-generated radio galaxies are processed through the RIME framework and then used as input for our neural network. The main neural network is the same as in \cite{Schmidt_2022} with small adjustments, such as twice the number of residual blocks. The training process runs smoothly, as shown in \autoref{fig:train_loss} and produces reconstructions that match the corresponding simulated sky simulation very well, as shown in \autoref{fig:compl_reco}.

To generalize the performance of the trained model, we evaluated the sources in image space using three methods that compare area, peak flux density, and integrated flux density. This was supplemented by a method adopted from the field of computer vision, which is called MS-SSIM. Each of these methods was applied to \num{10000} reconstructions from \texttt{radionets} and from \texttt{WSCLEAN}. The results are shown in \autoref{fig:hist-area} to \autoref{fig:hist-msssim}. While for the ratio of integrated flux densities, \texttt{radionets} and \texttt{WSCLEAN} perform equally well, \texttt{WSCLEAN} overestimates and underestimates for the areas and the MS-SSIM, respectively. The results for the peak flux densities for both methods vary widely due to resolution differences as shown in the comparison of the right images in \autoref{fig:image-pred} and \autoref{fig:reco_wsclean}. Overall we find that \texttt{radionets} is able to reconstruct far more complex source structures than in our previous paper, while only needing minor adjustments. Also, it performs at least as well as \texttt{WSCLEAN} for the main source properties. However, due to non-correctable beam-smearing effects and different effective resolutions, the comparison is not entirely fair.

Then, we divide up the results according to their SNR. The expectation is that for images with a high SNR the mean and standard deviations of our diagnostic methods are closer to their respective optimal values than for images with a lower SNR. This applies to both \texttt{radionets} and \texttt{WSCLEAN}. In \autoref{fig:mean-std-area} to \autoref{fig:mean-std-msssim}, one can see that this expectation holds in almost every case except for the peak flux densities, where the results are very close for every SNR besides 10. Thus, our network is robust and able to reconstruct complex structures despite a rather low SNR.

To summarize, we improved our simulations with two key features: (i) the GAN-generated radio galaxy images and (ii) the simulation of an interferometer measurement using RIME. With these two developments, we are able to train our network with more realistic data. As a result, the network is able to reconstruct the sources with similar accuracy compared to \texttt{WSCLEAN}, despite the increased complexity of the sources. Finally, the evaluation metrics suggest a better performance than the model in \cite{Schmidt_2022}.

Both frameworks, \texttt{pyvisgen} \citep{pyvisgen} for the simulation part and \texttt{radionets} \citep{radionets} for the model training, are available on GitHub. 

Finally, we give an outlook on future features of our deep learning-based approach.
We focus on the estimation of the reconstruction error, as outlined in \autoref{fig:cat}, by extending the neural network such that it computes uncertainties as well as visibility reconstructions.
The second goal is the application of our deep learning-based imaging approach to actual observation data.
In the future, we will focus on the minimization of data-simulation mismatches to train deep learning models suitable to reconstruct survey data of large parts of the sky.

\section{Acknowledgments}

MB acknowledges support from the Deutsche Forschungsgemeinschaft (DFG) under Germany’s Excellence Strategy -- EXC 2121 ”Quantum Universe” -- 390833306, funding through the DFG Research Unit FOR 5195 and and via the KISS consortium (05D23GU4, 13D22CH5) funded by the German Federal Ministry of Education and Research BMBF in the ErUM-Data action plan.

The authors acknowledge support from the project \enquote{Big Bang to Big Data (B3D)}. B3D is receiving funding from the program \enquote{Profilbildung 2020}, an initiative of the Ministry of Culture and Science of the State of North Rhine-Westphalia. The sole responsibility for the content of this publication lies with the authors.

DE acknowledges funding through the BMBF for the reasearch project D-MeerKAT-II and D-MeerKAT-III in the ErUM-Data context.

The authors acknowledge support from the German Science Foundation (DFG) within the Collaborative Research Center SFB1491.

We thank the anonymous referee for an insightful and stimulating report.

\bibliographystyle{aa}
\bibliography{references}

\FloatBarrier
\begin{appendix}
\section{Used software and packages}

\begin{minipage}{\textwidth}
In this work, we used \texttt{PyTorch} \citep{pytorch} as the fundamental deep learning framework. It was chosen because of its flexibility and the ability to directly develop algorithms in the programming language \texttt{Python} \citep{python}. In addition to \texttt{PyTorch}, we used the deep learning library \texttt{fast.ai} \citep{fastai} which supplies high-level components to build customized deep learning algorithms in a quick and efficient way.
For simulations and data analysis, the Python packages \texttt{NumPy} \citep{numpy}, \texttt{Astropy} \citep{astropy1, astropy2}, \texttt{Cartopy} \citep{cartopy}, \texttt{scikit-image} \citep{scikit-image}, and \texttt{Pandas} \citep{pandas} were used.
The illustration of the results was done using the plotting package \texttt{Matplotlib} \citep{matplotlib}.
A full list of the used packages and our developed open-source \texttt{radionets} framework can be found on github\footnote{\url{https://github.com/radionets-project/radionets}}.
More information about our developed open-source simulation framework \texttt{pyvisgen} can be found in the git repository\footnote{\url{https://github.com/radionets-project/pyvisgen}}.
\end{minipage}


\begin{figure*}[!htbp]
\centering
\begin{minipage}{\textwidth}
\section{Gaussian smoothing of GAN generated radio galaxies}
    \centering
    \includegraphics[width=0.5\hsize]{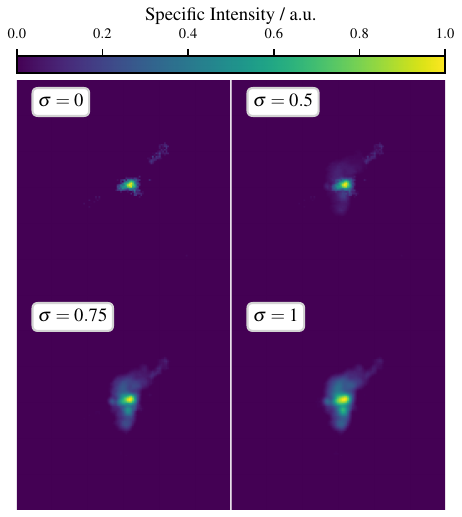}
    \caption{Effect of a Gaussian filter on the GAN-generated sources. We decided to use a smoothing of $\sigma=\SI{0.75}{\pixel}$ due to the best effect without creating false point sources as visible for $\sigma=\SI{1}{\pixel}$.}
    \label{fig:gaussian_filter}
\end{minipage}
\end{figure*}


\begin{figure*}
\centering
\begin{minipage}{\textwidth}
\section{Phase vs. imaginary part}
    \centering
    \includegraphics[width=\hsize]{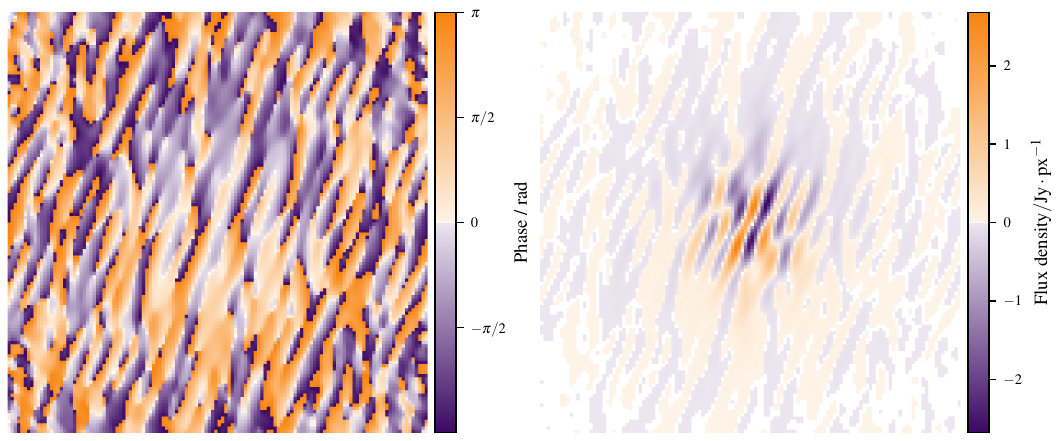}
    \caption{Fourier transformed source with the corresponding phase part (left) and imaginary part (right). The structures in the center of the images are very similar while the edges differ, especially in terms of intensity.}
    \label{fig:imag_vs_phase}
\end{minipage}
\end{figure*}

\FloatBarrier

\begin{table}
    \centering
\begin{minipage}{\textwidth}
\section{Computer setup}
    \centering
    \caption{Computer specifications for the setup used in the training process}
    \begin{tabular}{c|c|c}
    \toprule
    Part & Specification & Value \\
    \midrule
    GPU &  Nvidia DGX A100 & \SI{8}{\gigabyte}\\
    CPU & AMD Rome & 64 Cores @ \SI{3.35}{\giga\hertz}\\
    Hard drive & SSD & \SI{512}{\gigabyte} \\
    \bottomrule
    \end{tabular}
    \label{tab:computer-specifications}
\end{minipage}
\end{table}


\begin{table}[!htbp]
    \centering
\begin{minipage}{\textwidth}
\section{Antenna positions VLA B-configuration}
    \centering
    \caption{Overview of the antenna positions of the VLA used for the RIME in the Earth-centered coordinate system. The given positions correspond to the B-configuration of the VLA used for the FIRST observations.}
    \begin{tabular}{
    l
    S[table-format=-4.2]
    S[table-format=-4.2]
    S[table-format=-4.2]
    }
    \toprule
    Station Name &  {X / m} &  {Y / m} &  {Z / m} \\
    \midrule
    W32&1640.03 &   -4329.93    &   -2416.71\\
    N24&-1660.49&   -259.40     &   2454.42 \\
    W20&733.35  &   -1932.98    &   -1078.11\\
    E24&765.21  &   2889.45     &   -1108.88\\
    N32&-2629.09&   -410.65     &   3885.60\\
    E36&1534.56 &   5793.91     &   -2223.48\\
    N16&-801.40 &   -124.97     &   1182.12\\
    W8&152.76   &   -401.27     &   -223.40\\
    W28&1316.45 &   -3443.31    &   -1913.53\\
    W12&306.17  &   -804.58     &   -448.08\\
    W36&2000.07 &   -5299.80    &   -2962.89\\
    N28&-2091.44&   -326.59     &   3089.41\\
    E32&1253.25 &   4733.62     &   -1816.91\\
    N36&-3217.58&   -502.67     &   4756.12\\
    E28&998.67  &   3764.31     &   -1443.47\\
    N4&-74.82   &   -11.76      &   111.63\\
    E4&35.60    &   133.65      &   -51.10\\
    W4&46.92    &   -122.02     &   -67.61\\
    W16&499.84  &   -1318.00    &   -735.21\\
    N8&-243.60  &   -38.05      &   360.04\\
    E16&376.99  &   1440.99     &   -556.13\\
    E8&114.43   &   438.69      &   -169.49\\
    E20&560.10  &   2113.27     &   -810.70\\
    E12&229.47  &   879.60      &   -339.87\\
    N20&-1174.34&   -183.30     &   1734.24\\
    N12&-489.31 &   -76.31      &   721.52\\
    W24&1005.43 &   -2642.99    &   -1472.20\\
    \bottomrule
    \end{tabular}
    \label{tab:antennas-vla}
\end{minipage}
\end{table}

\FloatBarrier

\begin{figure*}[!htbp]
    \centering
    \begin{minipage}{\textwidth}
    \section{Visualize of Jones matrices}
    \centering
    \includegraphics[width=0.75\hsize]{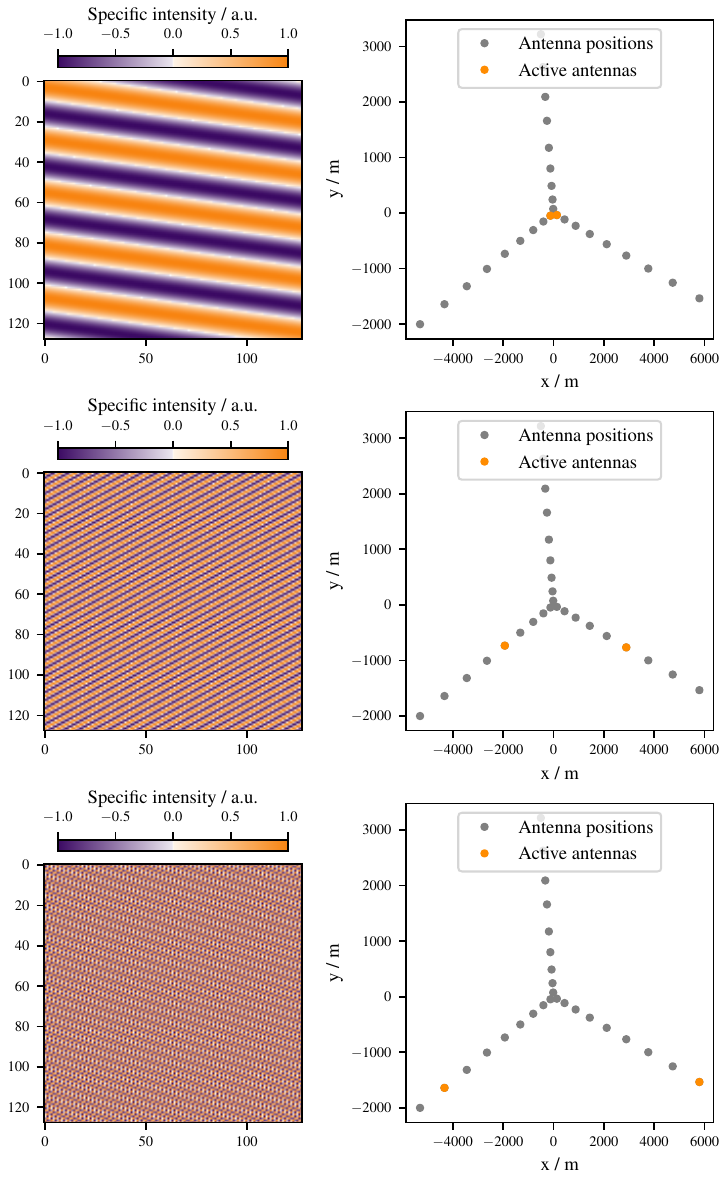}
    \caption{Visualization of Jones $\mathbf{K}(l, m)$, which represents the phase delay effect. Examples are given for three different antenna pairs of the VLA. With increasing baseline length, the fringe pattern becomes sharper, meaning that these baselines are more sensitive to small scale structures.}
    \label{fig:phase_delay}
    \end{minipage}
\end{figure*}

\begin{figure*}[!htbp]
    \centering
    \includegraphics[width=0.45\hsize]{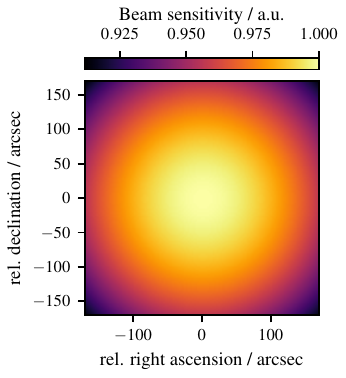}
    \caption{Visualization of Jones $\mathbf{E}(l, m)$, which represents the telescope's beam effects. In this work, the simulated field of views are small, which results in no significant beam sensitivity decrease towards the edges.}
    \label{fig:telescope_response}
\end{figure*}


\begin{table*}[!htbp]
    \centering
    \begin{minipage}{\textwidth}
    \section{Summary of evaluation methods results}
    \centering
    \caption{Results of the evaluation methods for data sets with different SNRs reconstructed with \texttt{radionets}}
    \begin{tabular}{c|c|c|c|c}
    \toprule
     SNR &  Area & Peak fluxes & Integrated fluxes & MS-SSIM \\
     \midrule
     0-10 & \num{0.945(146)} & \num{1.002(106)} & \num{0.936(107)} & \num{0.998(5)} \\
     10-20 & \num{0.950(95)} & \num{0.994(68)} & \num{0.958(57)} & \num{0.999(2)} \\
     20-30 & \num{0.974(85)} & \num{0.997(65)} & \num{0.978(40)} & \num{0.999(1)} \\
     30-40 & \num{0.980(72)} & \num{1.000(66)} & \num{0.988(29)} & \num{1.000(1)} \\
     40-50 & \num{0.981(70)} & \num{1.001(69)} & \num{0.990(30)} & \num{1.000(1)} \\
     50-60 & \num{0.987(75)} & \num{0.997(72)} & \num{0.990(28)} & \num{1.000(0)} \\
     $>60$ & \num{0.988(54)} & \num{1.000(68)} & \num{0.993(26)} & \num{1.000(0)} \\
     \bottomrule
    \end{tabular}
    \label{tab:snr}
    \end{minipage}
\end{table*}

\begin{table*}[!htbp]
    \centering
    \caption{Results of the evaluation methods for data sets with different SNRs reconstructed with \texttt{WSCLEAN}}
    \begin{tabular}{c|c|c|c|c}
    \toprule
     SNR &  Area & Peak fluxes & Integrated fluxes & MS-SSIM \\
     \midrule
     0-10 & \num{1.378(413)} & \num{0.479(68)} & \num{0.907(105)} & \num{0.822(84)} \\
     10-20 & \num{1.242(345)} & \num{0.468(57)} & \num{0.952(82)} & \num{0.881(54)} \\
     20-30 & \num{1.229(329)} & \num{0.481(58)} & \num{0.981(57)} & \num{0.919(41)} \\
     30-40 & \num{1.189(285)} & \num{0.488(58)} & \num{0.988(45)} & \num{0.938(33)} \\
     40-50 & \num{1.169(267)} & \num{0.499(57)} & \num{0.990(38)} & \num{0.949(30)} \\
     50-60 & \num{1.169(281)} & \num{0.493(55)} & \num{0.992(35)} & \num{0.955(27)} \\
     $>60$ & \num{1.063(196)} & \num{0.483(58)} & \num{1.008(38)} & \num{0.964(21)} \\
     \bottomrule
    \end{tabular}
    \label{tab:snr-wsclean}
\end{table*}

\FloatBarrier

\begin{figure*}[!htbp]
    \centering
    \begin{minipage}{\textwidth}
    \section{Signal to noise ratios}
    \centering
    \includegraphics[width=0.5\hsize]{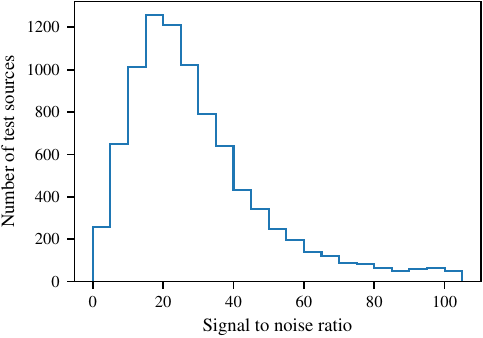}
    \caption{Overview of the SNRs of the sources inside the test data set. The majority of the sources have a SNR between 10 and 40.}
    \label{fig:hist_snr}
    \end{minipage}
\end{figure*}

\begin{figure*}[!htbp]
    \centering
    \begin{minipage}{\textwidth}
    \centering
    \includegraphics[width=\hsize]{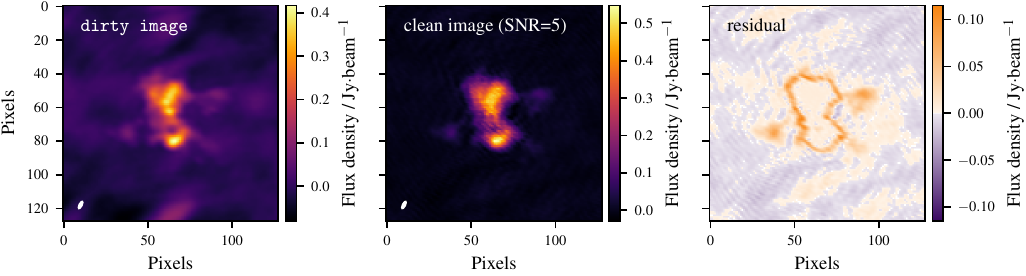}
    \includegraphics[width=\hsize]{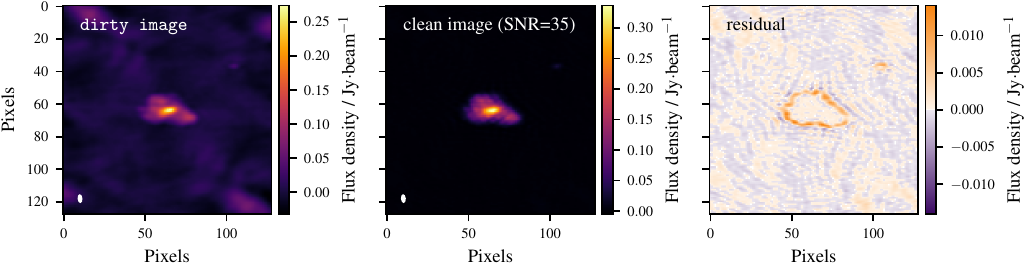}
    \includegraphics[width=\hsize]{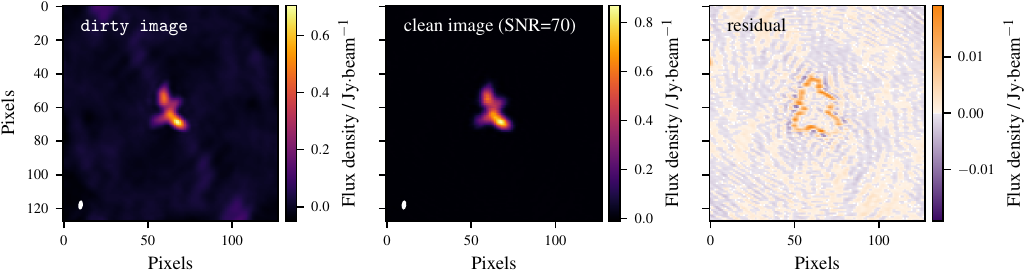}
    \caption{\texttt{WSCLEAN} results for simulations with different SNRs. Dirty image (left), clean image (middle), and remaining residual (right) are shown for a SNR of 5 (top), 35 (middle), and 70 (bottom).}
    \label{fig:vis_snr}
    \end{minipage}
\end{figure*}

\end{appendix}


\end{document}